**Ab initio investigation of the AlN:Er system**

Th. Pavloudis[1], V. Brien[2], J. Kioseoglou[1,*]

[1]Department of Physics, Aristotle University of Thessaloniki, GR-54124 Thessaloniki, Greece

[2]Institut Jean Lamour, UMR 7198, CNRS, Université de Lorraine, Boulevard des Aiguillettes, B.P. 239, 54506 Vandœuvre-lès-Nancy Cedex, France.

**Abstract**

In the present study an *ab initio* investigation on the AlN:Er system for concentrations of Er ranging from 0.78 to 12.5 % is presented. The crystallographic localisation of the rare earth atoms in the wurtzite lattice is determined, elucidating previously published experimental deductions, and the existence of a solid solution in the AlN:Er system in this range is confirmed. Er incorporation in the tetrahedral and octahedral insertion sites is shown to be thermodynamically metastable and is found to induce shallow states in the bandgaps. The effect of Er concentration on the lattice constants and bandgaps and bandstructures of the $Er_xAl_{1-x}N$ ternary compound is presented. Finally, in accordance with experimental specifications, Er incorporation in the AlNO system is also examined.

---

[*] Corresponding author. Tel: +302310 99 8312. E-mail address: sifisl@auth.gr

## 1. Introduction

A variety of lanthanide elements in III-nitride semiconductors exhibit significant room temperature luminescence [1,2,3,4]. Hence, rare-earth (RE) doped nitrides are very promising for a series of optoelectronic applications among them flat panel displays and high contrast thick dielectric electroluminescent display devices [1,2]. Although various studies on wurtzite nitride alloys have been reported [i.e. 5,6,7,8], the ultimate goal is to assess the theoretical interpretation of luminescent mechanisms of nitrides or oxynitrides doped with RE atoms.

The domain of application of such materials is wide, ranging from the telecoms or medicine to lighting, passing by solid-state lasers, optical displays or radiation and light detection. Lanthanide atoms on their own are not natural emitters but when properly inserted in semiconductors or specific insulators, the trivalent RE ions can be rendered optically active. The selection rules for the intra $4f$ energetic transitions, which are parity forbidden in the metallic state, can be relaxed due to crystal field effects and can lead the ions to produce sharp and strong luminescence peaks. As the symmetry of the site is ruling the way the electronic levels are split, the crystallographic localization of the RE ions can completely change the excitation-emission optical mechanism. The localization and the nature of the coordination spheres of the RE are indeed key aspects to understand and control the intensities of the optical centres emissions [9,10].

Several theoretical studies on the incorporation of RE elements in III-nitride alloys have been published. S. Petit et al [11], by the use of Density Functional Theory (DFT) calculations, concluded that Er atoms prefer to substitute Al atoms in wurtzite AlN and the resulting atomistic configuration is electrically active, introducing deep donor level in the bandgap. S. Sanna et al [12], also using DFT based calculations using the "frozen core" approach, in which the $4f$ electrons are included in the atomic core, investigated several atomistic configurations of the Er-O defect pairs in GaN. They concluded that the presence of oxygen does not affect significantly the atomic structure and the charge distribution around the Er atoms, while the Er-O defect pairs are thermodynamically stable. Er dopants in wurtzite GaN have been also onvistigated by the use of DFT-based tight-binding calculations [1,13]. In order to study substitutional defects and complexes with GaN native defects, $4f$ shell was explicitly included, considering an +U functional suitable for localized states [1,13]. Following this approach, Eu, Er and Tm impurities were found to prefer the Ga-sites in wurtzite GaN [14].

Along the quest for higher luminescent RE doped materials, researchers face the so-called "concentration quenching" phenomenon [3,15,16,17], observed when the doping content of the RE is pushed (several atomic %). The optical production drop is often justified by the precipitation of RE rich phases above a certain doping level, which vary from one matrix to another and promote the loss of energy due to $RE^{3+}$-$RE^{3+}$ electron transfers. The same "concentration quenching" was measured both in the infrared and visible domains [4]. In addition, Eu, Tm and Er implantation in AlN films showed that the wurtzite structure is preserved, while annealing recovers part of the produced lattice damage [18]. They concluded that the RE atoms occupy not only Al substitutional sites but also interstitial lattice sites [18].

Beyond the quest for higher optical efficiency of devices, the challenge is important to promote the understanding of the optical mechanisms [19]. The authors of this work intend to examine if theoretical calculations based on the electronic structure can confirm or not the diverse experimental observations. The attention is focused on the existence or not of a solid solution in the RE doping range of interest, attesting or not the Er rich phase precipitation for these ranges of compositions in the AlN:Er system. The work employs high accuracy *ab initio* calculations in an effort to examine the crystallographic localization of the Er atoms in the wurtzite cell and the structural and electronic effects of Er incorporation in AlN. Additionally, the localization and effect of Er in the AlNO:Er system, at O and Er concentrations as close as possible to experimentally realised samples, is also investigated.

## 2. Calculation methodology

The VASP *ab initio* simulation package [20] with Projector Augmented Wave (PAW) pseudopotentials [21,22] under the Perdew-Burke-Ernzerhof derivation of the Generalized Gradient Approximation (GGA-PBE) [23,24] is implemented in order to examine the incorporation of Er and O in AlN. A Gaussian smearing with a smearing width of 0.01 eV is employed for the calculation of the energies and the construction of the bandstructures. Γ-centered *k*-point meshes with a resolution of approximately 0.2 Å$^{-1}$ are used for the calculations, resulting in a 12x12x8 mesh for the primitive wurtzite cell and a 12x12x12 mesh for the primitive cubic cell. The meshes are scaled accordingly for the larger supercells. In all the relaxations the ions as well as the cell shape and volume are allowed to relax. The plane wave basis energy cutoff is 520 eV to compensate for the volume relaxations. Spin-polarized calculations are performed and the spin-resolved bandstructures are obtained for the models.

Supercells ranging from 32 to 256 atoms are constructed to study the localization of Er in AlN in doping ranges of 0-12.5 (1/128) atomic %, covering and going beyond the experimental doping range of 0-6 atomic % Er. Visualization of the atomic structures is performed with the VESTA visualization program [25].

For Al, N and O the standard pseudopotentials, which provide very high accuracy results, are used. The valencies of Al, N and O are 3, 5, and 6 respectively. Concerning Er, classic DFT does not handle well the $f$ electrons and provides an incorrect description of partially filled $f$ states due to self-interaction errors. The common way to handle the problem is to include $f$ electrons in the core electrons and a choice of pseudopotentials for most $f$ elements, where the number of $f$ electrons placed in the core varies and equals the total number of valence electrons minus the formal valency assumed by the element is provided. While this approach may produce good structural results, $f$ electrons are important for the electronic properties of the system and cannot be neglected. Therefore, an Er pseudopotential that includes 22 valence electrons in a $4f^{11}\ 5s^2\ 6s^2\ 5p^6\ 5d^1$ configuration with the remaining electrons included in the core is used in this work. An approach beyond classic DFT is needed to properly describe the behaviour of $f$ electrons, therefore the DFT+U method is employed. DFT+U is a method which introduces a strong intra-atomic interaction in a screened Hartree-Fock like manner as an on site replacement of DFT, and has been used in the past for RE nitrides [26].

In this work, the simplified, rotationally invariant, approach to DFT+U, introduced by Dudarev et al. [27] is used. In Dudarev's approach, the parameters $U$ and $J$, which control the effective on-site Coulomb interaction and the effective on-site exchange interaction respectively, are not entered separately and only the difference $U_{eff}=U-J$ is meaningful. Non-spherical contributions from the gradient corrections are included inside the PAW spheres, a factor essential for accurate total energies and band structure calculations for $f$-elements if the $+U$ approach is used.

$U_{eff}$ is considered as a free parameter and is optimized according to the procedure described at [2]. Bandstructure calculations of a 256-atom AlN supercell in which an Er atom replaces an Al atom are performed scaling $U_{eff}$ from 0 up to 10. For low $U_{eff}$ values, filled and empty levels related to the Er 4*f* electrons are located in the bandgap. These levels shift up towards the conduction band minimum (CBM) and down towards the valence band maximum (VBM) linearly with the $U_{eff}$ parameter and are finally ejected from the bandgap after a approximately $U_{eff}$~8.5 eV. The value of $U_{eff}$=8.6 eV, at which the 4*f* levels are located in the VBM and CBM and which is consistent with previously published results concerning ErN [26], is adopted and is used for all the calculations performed in this work. The +*U* potential is applied only to the *f* electrons of Er.

These calculation settings provide lattice constants of 4.842 Å for rocksalt ErN, within 0.1 % of the 4.839 Å experimental lattice constant [28], and predict a small indirect bandgap of 0.26 eV (direct bandgap at Γ equal to 3.42 eV). The Er-N bond length is 2.42 Å. The spin-resolved bandstructures of ErN are shown in Fig. 1. The optimal lattice constants for AlN are 3.128 Å and 5.019 Å with a direct bandgap of 4.06 eV.

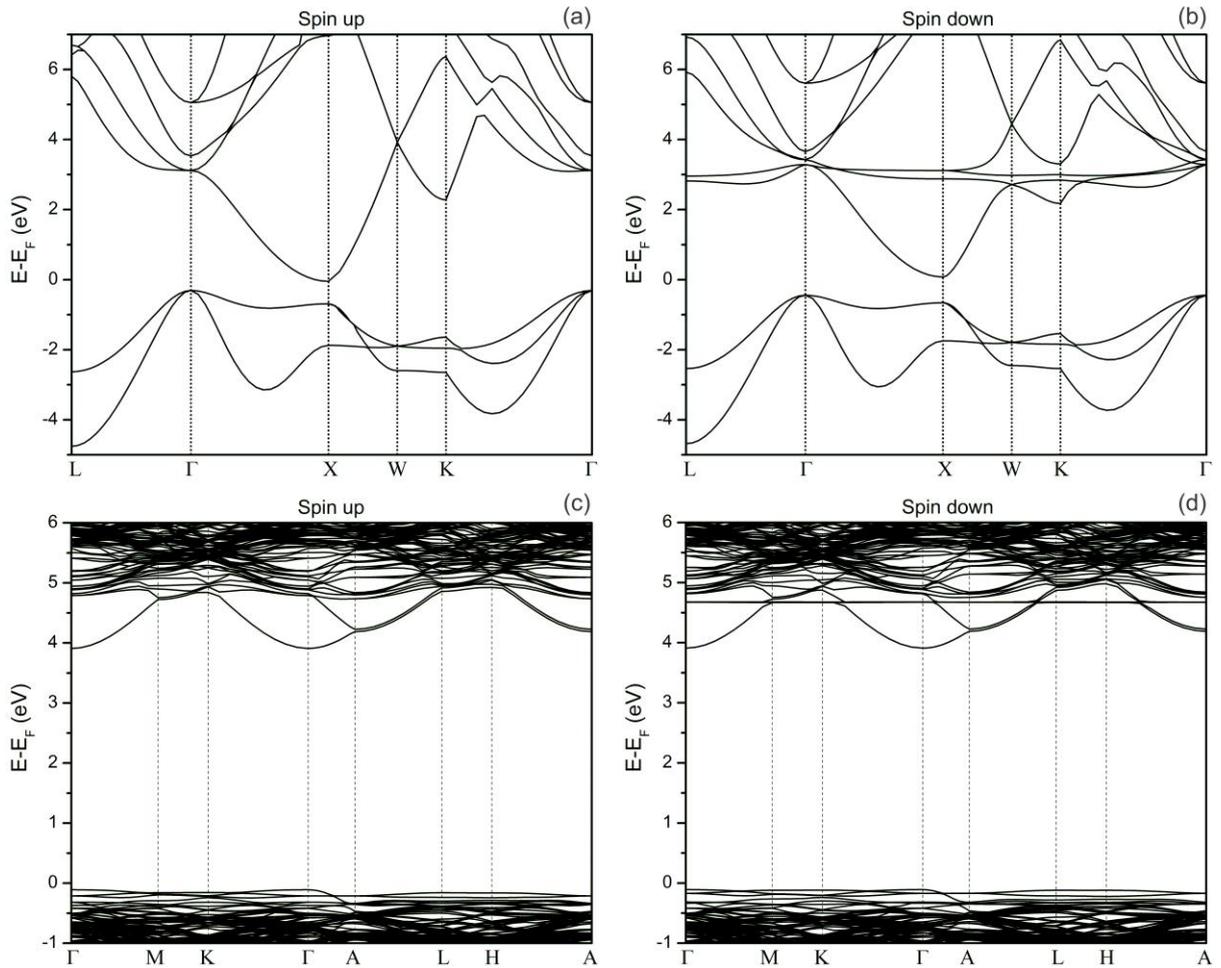

**Figure 1.** The spin-resolved bandstructures of (a)-(b) rocksalt ErN and (c)-(d) AlN:Er$_{Al}$ for $U_{eff}$ =8.6 eV. The Fermi level is at 0 eV.

## 3. Er incorporation in AlN

The atomistic configurations studied in the present work, are tailored according to the experimental elements published in Ref. [4]. The incorporation of Er in AlN is examined using one Er atom in 4x4x4 256-atom AlN supercells for a resulting Er concentration of approximately 0.78 % (1/128). The wurtzitic lattice of AlN intrinsically possesses three crystallographic sites for the RE atom. The three different localizations for Er that are considered are:

a) $Er_{Al}$: substituent Er in an Al atom position resulting in a $Er_{0.01}Al_{0.99}N$ alloy.
b) $Er_{i,O}$: interstitial Er in an octahedral position, i.e. at the 2a Wyckoff position (0, 0, z), shown in Figure 2(a)
c) $Er_{i,T}$: interstitial Er in a tetrahedral position which corresponds to the position in between the Al and N atoms, i.e. at the 2b Wyckoff position (1/3, 2/3, z), shown in Figure 2(b)

The defect formation energies $E_f[X_q]$ are calculated according to [29] by:

$$E_f[X_q] = E_{tot}[X_q] - E_{tot}[AlN, bulk] - \sum_i n_i \mu_i + q[E_{VBM} + E_F]$$

where $E_{tot}[X_q]$ and $E_{tot}[AlN, bulk]$ are the total energies after the relaxations of the defected supercells and the perfect AlN supercell, $n_i$ is the number of atoms $i$ added to the supercell ($n_i > 0$) of removed from the supercell ($n_i < 0$), $\mu_i$ are the corresponding chemical potentials of the species, $q$ is the charge state of the defect and $E_F$ is the is the Fermi level, referenced to the VBM of the bulk $E_{VBM}$. Al-rich conditions are assumed, i.e. the chemical potential of Al is $\mu_{Al} = \mu_{Al,bulk}$. Accordingly, the chemical potential of N is $\mu_N = E_{tot,AlN} - \mu_{Al,bulk}$ and finally the chemical potential of Er is $\mu_{Er} = E_{tot,ErN} - \mu_N$. The defect formation energies are calculated for the 0,+1,+2 and +3 charge states and are shown in Table 1.

**Table 1.** Defect formation energies of Er interstitials in AlN in the octahedral and tetrahedral positions under Al-rich conditions.

| Charge state | $Er_{i,O}$ | $Er_{i,T}$ |
|---|---|---|
| 0 | 11.92 | 14.14 |
| +1 | 8.31 | 10.22 |
| +2 | 4.33 | 6.77 |
| +3 | 0.62 | 3.86 |

Between the two interstitial positions, the Er atom is found to clearly prefer the octahedral one for all charge states. The defect formation energy for $Er_{i,O}$ is 11.92 eV for the uncharged case and only 0.62 eV for the +3 charged case. Comparably, the defect formation energy for $Er_{i,T}$ is 14.14 eV for the uncharged case and 3.86 eV for the +3 charged case. The defect formation energy for $Er_{Al}$ is s a flat 1.23 eV and this is deemed the globally energetically favourable case. The Er atom in this case is checked for + and – charge states and is preferably not charged, existing in its trivalent oxidation state. However, Er insertion in the interstitial sites, especially the octahedral site, is not unlikely, especially taking into account the thermodynamics of the growth process.

These calculations bring new information: they suggest the existence of a solid solution in AlN:Er for low contents of Er and prove that Er solubility in AlN is not pushed by the growth process. Er clustering in AlN is examined by adding an additional Er atom the 256-atom supercell in distinct 2 positions: next to the first Er atom and directly below the first atom at a distance equal to $2c_{AlN}$. The distances after the relaxations between the two Er atoms are 3.27 Å and 10.05 Å for the two cases respectively. The formation energy difference between the two cases is found to be equal to 0.14 eV in favour of the second case, proving that Er clustering in AlN is not energetically preferable.

The demonstrated preference of Er atoms for Al sites is in agreement with the experimental results obtained in Ref. [4]. The evolution of the relative intensities of the different wurtzite XRD peaks as a function of Er doping could not be interpreted only with the substitution of Al by Er atoms and had to take in account additional insertion sites. The authors consequently believe that the occupation of insertion sites (mainly the octahedral sites) deduced in [4] result from the extra energy brought by the experimental growth process itself. Indeed, magnetron reactive sputtering is a process able to lead to metastable phases and is notably exploited to stabilize new phases that may be thermodynamically metastable.

The effect on the lattice for the first case, where Er substitutes an Al atom, is small and concerns mainly the Al-N bond lengths and angles between the Al atoms in the neighbourhood of the Er atom and the 4 N atoms that are bonded with the Er atom. The Er-N basal and axial bond lengths are 2.14 Å and 2.18 Å respectively. The distortion in the lattice is noticeable in Figures 2(a) and 2(b) for the octahedral and tetrahedral interstitials respectively, where the neighbouring to the Er atom Al and N atoms are significantly displaced. The distance between the Al and N atoms, in between which the Er atom lies in the tetrahedral position, is bigger by 1.30 Å for uncharged and 1.33 Å for the +3 charged defect compared to the respective distance in bulk AlN. The AlN atoms that are bound to the displaced atoms are also displaced. The lattice distortion is intense for the two layers above and below the Er atom, after which AlN becomes bulk-like. The distances of the Er atom from the surrounding 4 N atoms are much smaller for the +3 case, dropping to an average of 2.14 Å compared to the average of 2.21 Å for the uncharged defect.

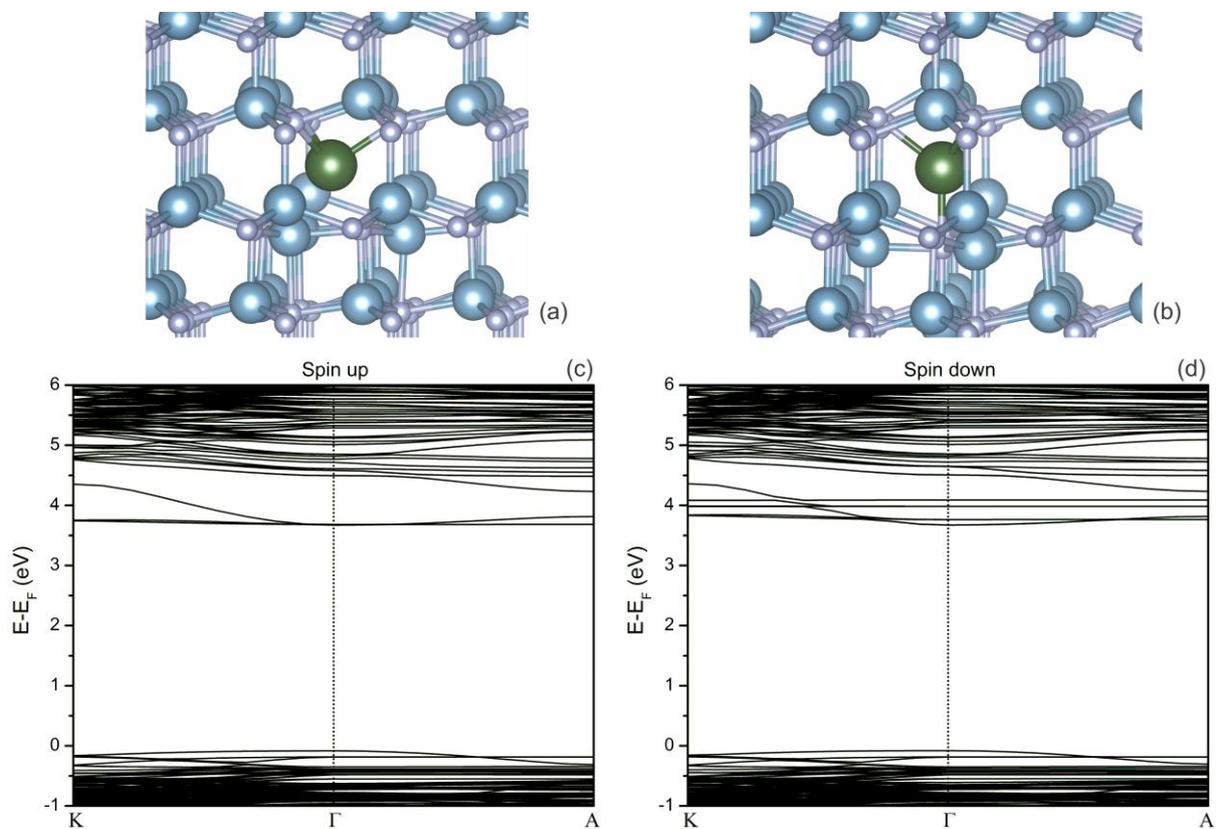

**Figure 2.** The (a) $Er_{i,O}$ and (b) $Er_{i,T}$ localizations in the AlN wurtzite structure after the relaxations for a +3 charge state. Small silver spheres represent N atoms, big blue spheres Al atoms, and the biggest green spheres correspond to Er atoms. (c) and (d) The spin-resolved bandstructure of AlN:$Er_{i,O}$. The defect charge is +3.

The Er atom is better accommodated in the octahedral position, where there is more space compared to the tetrahedral position, hence the lower energy of the octahedral interstitial. Starting from the uncharged case, where the lattice distortions in the Er-surrounding atoms are approximately 0.25 Å, the Er atom gradually pushes the two Al atoms directly beneath it for a final displacement of 1.00 Å. The Er atom distances from the 3 N atoms above the Er atom follow the same pattern as in the tetrahedral position, being gradually reduced from 2.17 Å for the uncharged defect to 2.14 Å for the +3 charged defect.

The lattice distortion and the bonding of Er with the neighbouring N atoms has a significant effect on the electronic properties of the system for both cases. Although there is a known large difference between the calculated and experimental bandgaps, a well-known issue for DFT calculations, qualitative results can be deduced from the bandstructure plots. When Er replaces an Al atom the bandgap is smaller, calculated at 4.02 eV. The bandgap, as mentioned before, is free of any deep or shallow states and is direct. That is not the case when Er is inserted as an interstitial: shallow partially occupied states are induced in the bandgap close to the CBM for both the tetrahedral and octahedral interstitials. The behaviour is more pronounced for the tetrahedral interstitial with more and deeper levels, up to 1eV lower than the CBM, are introduced in the bandgap. The spin-resolved bandstructure for the octahedral interstitial is shown in Figures 2(c) and (d).

## 4. Investigation of the $Er_xAl_{1-x}N$ ternary alloy

Following the aforementioned results, the focus is shifted to the $Er_xAl_{1-x}N$ ternary compound and the preferable positions of Er in AlN for more cases with different Er concentrations are examined. Specifically, two 2x2x2 32-atom supercells are used where one and two Er atoms replace Al atoms for an Er content of 6.25 % (1/16) and 12.5 % (2/16). An additional 3x3x3 108-atom supercell is constructed in order to examine the Er concentration of 1.85 % (1/54). The results include the previously mentioned 4x4x4 256-atom supercell with an Er concentration of approximately 0.78 % (1/128). All possible configurations are checked for a 2x2x2 32-atom supercell with the two Er atoms in a detailed examination that leads to the energetically preferable configuration, which is found to be an ordered structure in accordance with the results concerning ternary nitride compounds presented in Ref. [5].

Regarding the structural results, the Er-N basal and axial bond lengths are almost constant for all the models, independent of the the Er content, at approximately 2.14-2.15 Å and 2.18-2.19 Å respectively, with the exception of the 12.5 % case where the axial bond length becomes 2.25 Å. The relaxed lattice constants of the models and the bandgaps obtained from the spin-up bandstructure plots as a function of the Er content are presented in Figures 3(a) and 4(b) respectively. The lattice constants and the bandgaps follow harmonic relations that can be described by the quadratic Vegard's law. The law is formulated as:

$$a = x \cdot a^{AlN} + (1-x) \cdot a^{ErN} - b_\alpha \cdot x \cdot (1-x)$$

and

$$c = x \cdot c^{AlN} + (1-x) \cdot c^{ErN} - b_c \cdot x \cdot (1-x)$$

for the lattice constants and as

$$E_g = x \cdot E_g^{AlN} + (1-x) \cdot E_g^{ErN} - b_{E_g} \cdot x \cdot (1-x)$$

for the bandgaps.

The corresponding fits to the data yield the bowing parameters $b_\alpha = 0.64$ Å, $b_c = -3.50$ Å and $b_{E_g} = 28.01$ eV. The large bowing parameter for the bandgaps is worth noting. The bandgaps drop from 4.06 eV for AlN, to 4.02 eV for an Er concentration of 0.78 %, 3.92 eV (direct bandgap 3.97 eV) for an Er concentration of 1.85 %, 3.47 eV (direct bandgap 3.74 eV) for a concentration of 6.25 % and 3.18 eV (direct bandgap 3.49 eV) for a concentration of 12.5 %. The bowing parameter for the direct bandgaps at Γ is much smaller and equal to 7.21 eV. If a crude underestimation of the bandgap by a factor 1.5 is considered, which is approximately true for DFT and holds for AlN in particular, the bandgap bowing parameter becomes 41.57 eV.

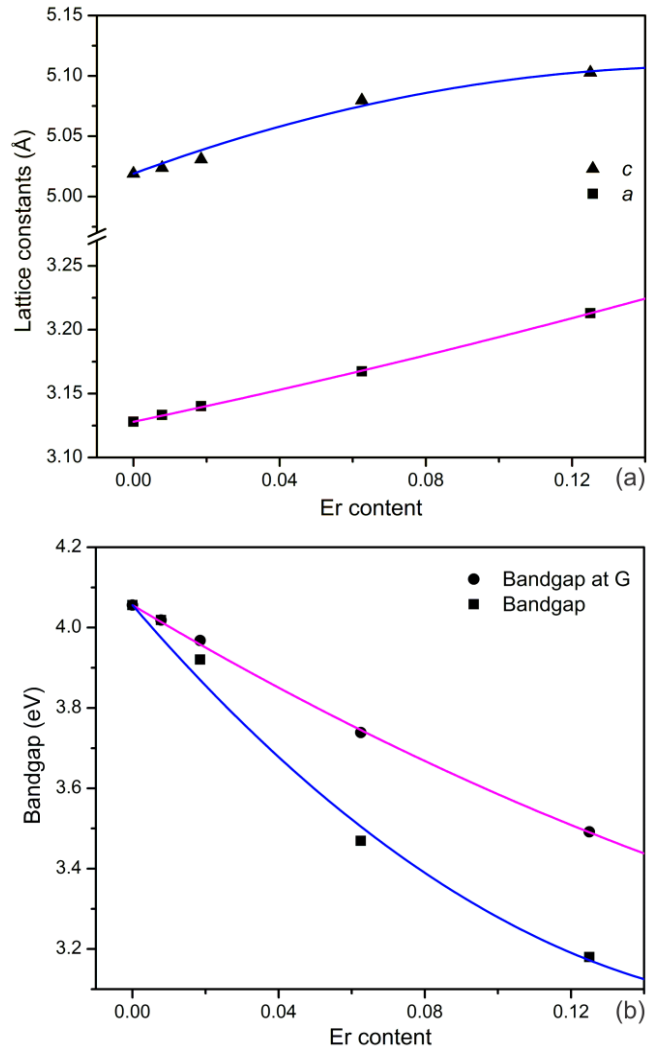

**Figure 3.** (a) The *a* (rectangles) and *c* (triangles) lattice constants of the Er$_x$Al$_{1-x}$N alloys as a function of the Er content and (b) the bandgaps of the Er$_x$Al$_{1-x}$N alloys as a function of the Er content. The blue and magenta curves are a second order polynomial fit according to the quadratic Vegard's law.

The calculations also bring to light the change of nature of the bandgap of the ternary compound, which is direct only at 0.78 % and switches to indirect for larger concentrations. The VBM is shifted to the L high-symmetry point for Er concentrations from 0.78 % to 6.25 % and to the M high-symmetry point for an Er concentration of 12.5 %. This behaviour will have consequences on the luminescence mechanisms.

It should be noted that the hexagonal structure of ErN is also examined. Relaxations starting from a wurtzite structure end in a hexagonal structure with *a* = 3.940 Å, *c* = 4.919 Å and *u* = 0.5 that belongs to space group 194. The relatively small $c_{ErN}$, even smaller than $c_{AlN}$, explains the large $b_c$. The formation energy difference between hexagonal and rocksalt ErN is 0.97 eV per Er-N pair. A linear interpolation between the formation energies of hexagonal AlN and ErN and rocksalt AlN and ErN shows that the hexagonal structure is dominant for an Er content lower than 26 %. After this point, the rocksalt structure is energetically preferable for the ternary alloy. The investigations performed in this paper concerning wurtzite $Er_xAl_{1-x}N$ are well within this limit. For example, the formation energy difference of wurtzite and rocksalt $Er_{0.125}Al_{0.875}N$ is found to be 0.44 eV per X-N pair in favour of the wurtzite phase.

An estimation of the average distance $r$ between two Er atoms is performed. Assuming AlN:Er is a solid solution, one can suppose a total dilution of Er. The lattice parameters of the supercells that correspond to one Er atom are deducted from the lattice parameters of the unit cells extracted by XRD for samples with known Er concentrations. The theoretical lattice parameters are readily obtained from the DFT relaxation results. The volume of the supercells containing one Er atom is calculated taking into account the hexagonal shape of the cell. The average distance $r$ between two Er atoms is then considered as the diameter of the sphere that corresponds to this volume. In Figure 4 the theoretical and experimental average distances $r$ between two Er atoms are presented as a function of the Er concentration.

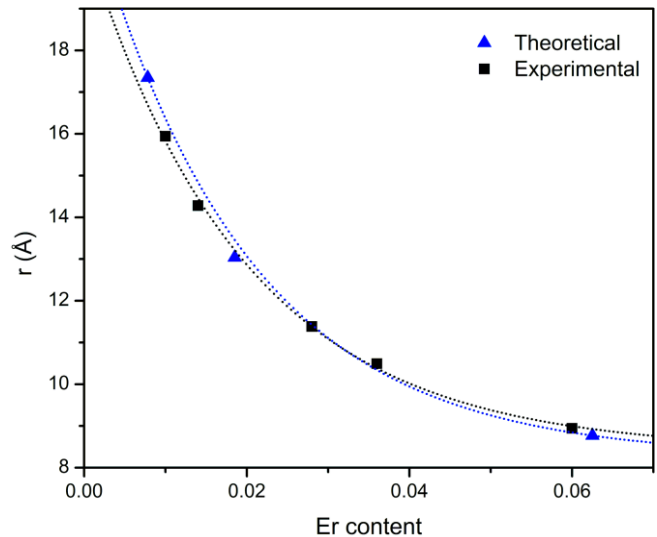

**Figure 4.** The average distance $r$ between two Er atoms calculated using experimental and theoretical results as a function of the Er content. The corresponding dotted curves are exponential decay fits on the two sets of data.

There is a very good agreement between experimental and DFT results. The curves converge to a plateau for high Er concentrations, which is calculated at 8.42 Å for the experimental and 8.24 Å for the theoretical results. The Er-Er distance for concentrations lower than 1 %, which is equivalent to the optimal industrial concentration, is larger than 15 Å. The Er atoms reside far away from each other and are not correlated in any way. A conclusion can be made that no Er-Er interactions will be noticed in the samples. The agreement of the theoretical results by the experimental measurements confirm that Er ions in AlN:Er exhibit a dilute and volumic distribution for a rather wide range of Er concentrations.

The DFT calculations presented in this work comfort that the classical behaviour of concentration quenching measured on AlN:Er films prepared by magnetron sputtering in the infra-red and visible domains cannot be justified by the precipitation of Er rich phases. Results suggest that, unlike in some other systems, the quenching is not due to the fact that the REs reach their solubility limit [30]. The theoretical work brings a theoretical justification to the results obtained after the experimental investigations done with XRD in Brien and Boulet [4].

## 5. Investigation of $Er_xAl_{1-x}N_yO_{1-y}$

As a final piece to the puzzle, the $Er_xAl_{1-x}N_yO_{1-y}$ alloy is examined. The incorporation of O in AlN is known to occur with the replacement of an N atom by an O atom in the AlN wurtzitic lattice [31]. A charge of +1 per O atom is considered in the calculations according to Refs. [29,32].

In the first case a 256-atom supercell in which an O atom has replaced an N atom is used, resulting in an O concentration of 0.78 %. In the second case a 2x2x2 32-atom supercell in which an O atom has replaced an N atom is multiplied by 2 along all directions, ending up in a 4x4x4 256-atom supercell with 8 O atoms, for an O concentration of 6.25 %, in an effort to mimic the O content of the samples mentioned in Ref. [4]. Consequently, an Er atom is placed on all possible Al positions and the resulting supercells are relaxed again.

In all cases, the energetically favourable positions by a margin of several eV for Er as a substituent are the Al positions that are closer to the O atoms. When situated at such a position, the Er atom creates bonds with one O atom and three N atoms. A simple explanation of this behaviour is that Er, having almost 2 times the size of Al, is better accommodated near the O atom, where more space is created because of its smaller size compared to the size of the N atoms. There is a slight difference between the possible positions next to the O atoms. The hexagonal symmetry of the cell imposes two inequivalent places for the Er atom when bonded to the O atom, one resulting in a basal configuration and resulting in axial configuration [12].

In agreement with the literature, the basal configuration is found to be energetically preferable, by a margin of 0.22 eV for an O content of 0.78 %. This behavior is more pronounced when the O content is higher: the formation energy difference for an O concentration of 6.25 % is 0.33 eV. This can be explained due to the proximity of the Er atom to a second O atom along the basal plane. The increased O concentration has only a marginal effect on the lattice. The Er-O bond has a length of 2.18 Å for the basal and 2.21-2.22 Å for the axial configurations for both 0.78 % and 6.25 % O. The energetically preferable configuration of $Er_{0.01}Al_{0.99}N_{0.94}O_{0.06}$ is shown in Figure 5(a) and the corresponding spin-resolved bandstructure is shown in Figure 5(b) and (c). The Er-O interaction is found to alter the bandstructure. A shallow level is introduced by the O atoms, in agreement with the literature. The level is moved towards the CBM by 0.01 eV for 0.78 % and by 0.03 eV for 6.25 % O by the Er presence. States are induced by the Er atom in the spin-down bandstructure near the CBM.

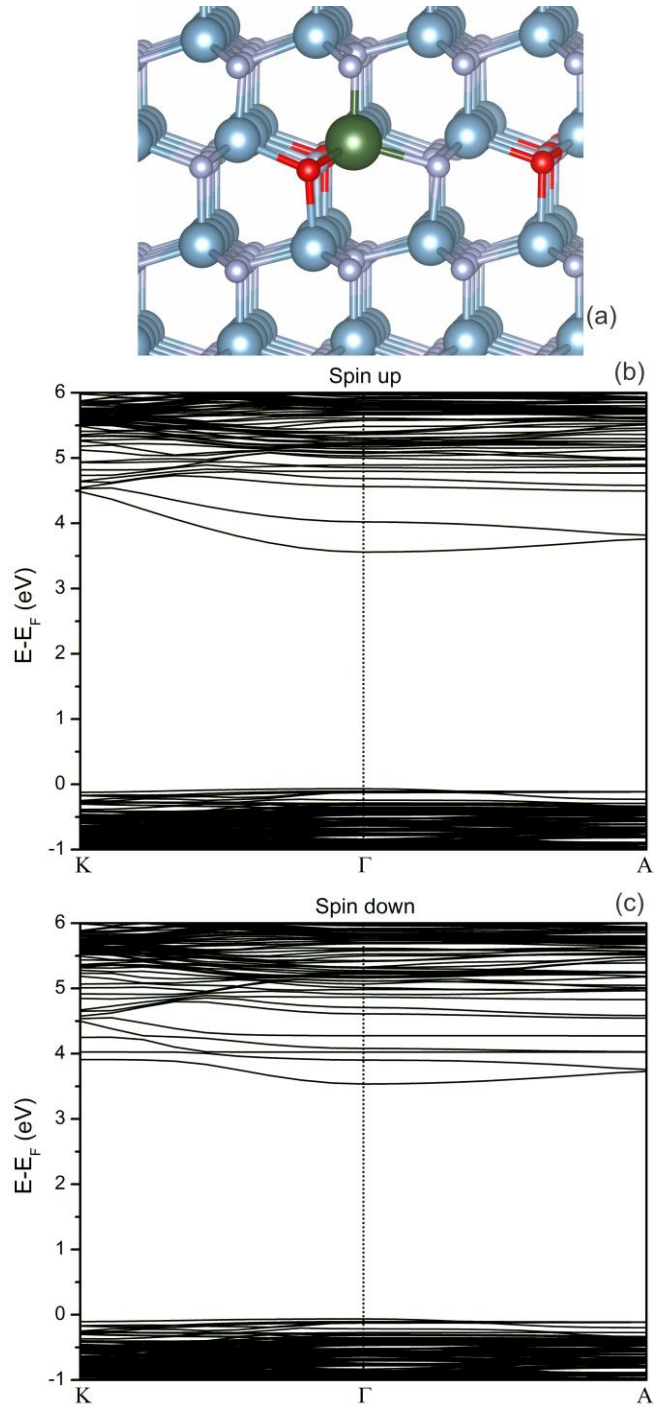

**Figure 5.** (a) The preferable configuration for $Er_{0.01}Al_{0.99}N_{0.94}O_{0.06}$. Small silver spheres represent N atoms, small red spheres O atoms, big blue spheres Al atoms, and the biggest green spheres correspond to Er atoms. (b)-(c) the corresponding spin-resolved bandstructures. A charge of +1 per O atom is considered.

## 5. Conclusions

In the present study an in depth analysis based on experimental results and *ab initio* calculations is performed on the structural and electronic properties of the AlN:Er system for low to moderate Er concentrations. The possible structural positions of Er atoms are examined as well as their influence of the bandstructure. In addition, the localisation of O atoms in the $Er_xAl_{1-x}N$ alloy is elucidated taking into account all possible atomistic configurations.

The absence of extra phases previously experimentally observed by dark field STEM images or electron and X-ray diffraction data in [4] is confirmed theoretically for the studied range, expanded even further by the DFT up to 12.5 %. The DFT+U study confirms Er is a full part of the wurtzite hexagonal cell; and confirms the existence of the $Er_xAl_{1-x}N$ solid solution on the quoted domain.

The lanthanide atom shows a preference for a regular metallic atom position as a substituent in the nitride but may also populate the octahedral and tetrahedral insertion sites of the hexagonal lattice, with a preference to the former ones between the two. Er incorporation in these insertion sites is metastable and pushed by the out-of-balance sputtering process and induces shallow states in the bandgap.

The bandgap of $Er_xAl_{1-x}N$ is shown to switch from a direct one to an indirect one at values of Er doping corresponding to the maximum efficiency of radiative emissions. In addition, the unintentional O doping of $Er_xAl_{1-x}N$ alloy is found to affect the electronic properties of the system.

The industrial potentialities of AlN:Er materials prepared by sputtering are consolidated, as their temporal stability is assessed thanks to the stability of solid solutions.

**Acknowledgements**
This work was supported by computational resources granted from the Greek Research & Technology Network (GRNET) in the National HPC facility 'ARIS' under the project under the project ATON (ID pr002004).

# References

[1] A. J. Steckl, J. C. Heikenfeld, D. S. Lee, M. J. Garter, C. C. Baker, Y. Wang, and R. Jones, IEEE J. Select. Top. Quantum Electron. 8, 749 (2002). doi: 10.1109/JSTQE.2002.801690.

[2] B. Hourahine, S. Sanna, B. Aradi, C. Kohler, and T. Frauenheim, Physica B 376–377, 512–515,(2006). doi: 10.1016/j.physb.2005.12.130

[3] H. Rinnert, S.S. Hussain, V. Brien, J. Legrand, P. Pigeat, Journal of Luminescence. 132 (2012) 2367–2370. doi:10.1016/j.jlumin.2012.04.008.

[4] V. Brien, P. Boulet, Acta Materialia. 90 (2015) 37–45. doi:10.1016/j.actamat.2015.02.022.

[5] Th. Pavloudis, J. Kioseoglou, T. Karakostas, P. Komninou, Computational Materials Science. 118 (2016) 22–31. doi:10.1016/j.commatsci.2016.02.036.

[6] M.J. Winiarski, Computational Materials Science, 108 (2015) 14-16. doi: 10.1016/j.commatsci.2015.06.013

[7] F. Tasnádi, F. Wang, M. Odén, I.A. Abrikosov, Computational Materials Science, 103 (2015) 194-199. doi: 10.1016/j.commatsci.2015.03.030.

[8] J. Guerrero-Sánchez, G.H. Cocoletzi, J.F. Rivas-Silva, N. Takeuchi, Computational Materials Science, 106 (2015) 155-160. doi: 10.1016/j.commatsci.2015.04.050

[9] A.J. Kenyon, Progress in Quantum Electronics. 26 (2002) 225–284. doi:10.1016/S0079-6727(02)00014-9.

[10] G.H. Dieke, R.A. Satten, American Journal of Physics. 38 (1970) 399–400. doi:10.1119/1.1976350.

[11] S. Petit, R. Jones, M. J. Shaw, P. R. Briddon, B. Hourahine, and T. Frauenheim Phys. Rev. B 72, 073205 (2005). doi: 10.1103/PhysRevB.72.073205

[12] S. Sanna, U. Gerstmann, W. Schmidt, Mater. Res. Soc. Symp. Proc., 1342, 93, (2011). doi: 10.1557/opl.2011.1050

[13] S. Sanna, B. Hourahine, U. Gerstmann, and Th. Frauenheim Phys. Rev. B 76, 155128 (2007). doi: 10.1103/PhysRevB.76.155128

[14] S. Sanna, B. Hourahine, Th. Frauenheim, and U. Gerstmann, phys. stat. sol. (c) 5, No. 6, 2358–2360 (2008). doi:10.1002/pssc.200778667

[15] P.G. Kik, A. Polman, Materials Science and Engineering: B. 81 (2001) 3–8. doi:10.1016/S0921-5107(00)00667-X.

[16] S. Chen, A. Uedono, J. Seo, J. Sawahata, K. Akimoto, Journal of Crystal Growth. 311 (2009) 3097–3099. doi:10.1016/j.jcrysgro.2009.01.048.

[17] F. Benz, J.A. Guerra, Y. Weng, A.R. Zanatta, R. Weingärtner, H.P. Strunk, Journal of Luminescence. 137 (2013) 73–76. doi:10.1016/j.jlumin.2012.12.028.

[18] K. Lorenz, E. Alves, T. Monteiro, M.J. Soares, M. Peres, P.J.M. Smulders, Nuclear Instruments and Methods in Physics Research Section B: Beam Interactions with Materials and Atoms, 242, 307-310, (2006). doi: 10.1016/j.nimb.2005.08.037

[19] Z. Dridi, A. Lazreg, H. Rozale, B. Bouhafs, Computational Materials Science, 48 (4) (2010) 743-748. doi: 10.1016/j.commatsci.2010.03.025.

[20] G. Kresse, J. Hafner, Phys. Rev. B. 47 (1993) 558–561. doi:10.1103/PhysRevB.47.558.

[21] P.E. Blöchl, Phys. Rev. B. 50 (1994) 17953–17979. doi:10.1103/PhysRevB.50.17953.

[22] G. Kresse, D. Joubert, Phys. Rev. B. 59 (1999) 1758–1775. doi:10.1103/PhysRevB.59.1758.

[23] J.P. Perdew, K. Burke, M. Ernzerhof, Phys. Rev. Lett. 77 (1996) 3865–3868. doi:10.1103/PhysRevLett.77.3865

[24] J.P. Perdew, K. Burke, M. Ernzerhof, Phys. Rev. Lett. 78 (1997) 1396–1396. doi:10.1103/PhysRevLett.78.1396

[25] K. Momma, F. Izumi, Journal of Applied Crystallography. 44 (2011) 1272–1276. doi:10.1107/S0021889811038970

[26] P. Larson, Walter R. L. Lambrecht, A Chantis, M. van Schilfgaarde, Phys. Rev. B 75 (2007) 045114. doi: 10.1103/PhysRevB.75.045114

[27] S. L. Dudarev, G. A. Botton, S. Y. Savrasov, C. J. Humphreys, A. P. Sutton, Phys. Rev. B. 57 (1998) 1505-1509. doi: 10.1103/PhysRevB.57.1505

[28] R.W.G. Wyckoff, Crystal Structures Krieger, Malabar, FL, (1982), Vol. 1.

[29] Q. Yan, A. Janotti, M. Scheffler, C.G. Van de Walle, Appl. Phys. Lett 105 (2014) 111104. doi: 10.1063/1.4895786

[30] A. Polman, Journal of Applied Physics. 82 (1997) 1–39. doi:10.1063/1.366265.

[31] C.G. Van de Walle, J. Neugebauer, Journal of Applied Physics. 95 (2004) 3851–3879. doi:10.1063/1.1682673.

[32] N.J. Cherepy, S.A. Payne, N.M. Harvey, D. Åberg, Z.M. Seeley, K.S. Holliday, I.C. Tran, F. Zhou, H.P. Martinez, J.M. Demeyer, A.D. Drobshoff, A.M. Srivastava, S.J. Camardello, H.A. Comanzo, D.L. Schlagel, T.A. Lograsso, Optical Materials, 54, 14-21, ( 2016). doi:10.1016/j.optmat.2016.02.008